\documentstyle[12pt]{article}

\begin{document}
\vspace{0.5in}
\begin{center}
\vspace{0.5in}
{\Large \bf An Estimate of the Proton Singlet Axial Constant}
\vspace{0.5in}

{\large \bf G.T.~Gabadadze}\\ \vspace{0.1in}

{\it Institute for Nuclear Research of the Russian Academy of
Sciences,
\\117312 Moscow, Russia}

{\it Bogoliubov Laboratory of Theoretical Physics, Joint Institute
for Nuclear Research, \\141980 Dubna, Russia }

\end{center} \vspace{0.1in}
\begin{center} {\bf Abstract} \end{center}

The value of the proton singlet axial constant is estimated.
It has been shown that the axial anomaly plays a crucial role in this
calculation. Obtained result is sufficiently suppressed in comparison with
the naively expected one. The magnitude of the strange quark contribution
for the proton state is also computed approximately.

\vspace{0.5in}

In the last time the problem of calculation of the nucleon singlet
axial constant has intensively been discussed in the literature.  On
the classical level, this constant determines the value of spin
carried by the valence quarks inside the nonrelativistic nucleon.
However, recent experimental data \cite {EMC} lead one to the quantity
which is much smaller than the naive theoretical value.  The
authors of ref. \cite {ET} (see also \cite {ARCCM}) have firstly
pointed out that it is the axial anomaly which suffers the naive
interpretation at the quantum level. As a result, the apparent  "spin
crisis" was resolved \cite {Venezia}.

In spite of this, the problem of calculation of the
nucleon singlet axial constant is far from being solved successfully.
Based on the
detailed analysis \cite {5}-\cite {10}, one can conclude that both
perturbative and nonperturbative contributions are essential for this
calculation.

In the present paper, we will try to estimate the value of the proton
singlet axial constant taking into account characteristic
peculiarities of the singlet axial channel.

Essentially, the problem is as follows: experimental
measurements \cite {EMC} have allowed one to calculate the integral
over the Bjorken
variable of the first structure function of polarized deep-inelastic
lepton-nucleon scattering \cite {EMC}. In accordance with the Ellis-Jaffe sum
rule \cite {EJ}, this integral is related to the axial constants for
the corresponding nucleon
$$
\int_0^1
g_1^p(x,Q^2)dx=C^{NS}(1,\alpha_s(Q^2))[{1\over12}g_A^{(3)}+
{1\over36}g_A^{(8)}]+
$$
$$
+ C^{S}(1,\alpha_s(Q^2)){1\over9}\tilde g_A^{(0)}. \eqno (1)
$$
Here $C^{NS}(1,\alpha_s(Q^2))$  and $C^{S}(1,\alpha_s(Q^2))$ are the
nonsinglet and  singlet operator co\-effi\-ci\-ent functions
normalized to unity at the tree level. The perturbative expansion for
the nonsinglet coefficient function is known at the
next-next-to-leading approximation of perturbation theory \cite
{NSCF} whenever it is known at the next-to-leading order for the
singlet one  \cite {SCF}.  The  axial constants in eq. (1) are
normalized as follows:

$$
<P,S|A_\mu^{(3)}|P,S>=<P,S|{\bar{u}}\gamma_{\mu}\gamma_5u
-{\bar{d}}\gamma_{\mu} \gamma_5d|P,S>=g_A^{(3)} S_{\mu}, \eqno (2a)
$$

$$
<P,S|A_\mu^{(8)}|P,S>=<P,S|{\bar{u}}\gamma_{\mu}\gamma_5u+{\bar{d}}
\gamma_{\mu}\gamma_5d-2{\bar{s}}\gamma_{\mu}\gamma_5s|P,S>=
g_A^{(8)}S_{\mu}, \eqno (2b)
$$

$$
<P,S|A_\mu^{(0)}|P,S>=<P,S|{\bar{u}}\gamma_{\mu}\gamma_5u+{\bar{d}}
\gamma_{\mu}\gamma_5d+{\bar{s}}\gamma_{\mu}\gamma_5s|P,S>
= g_A^{(0)} S_{\mu}, \eqno (2c)
$$
with $S_\mu$ being the proton spin four-vector .
Here the state vectors  are normalized in the Fock space in
accordance with the following condition: $$ <P(k_1)|P(k_2)>=(2\pi)^3 2
k_1 ^0 {\delta}^{(3)}(\vec{k}_2-\vec{k}_1).
$$ The renormalization group invariant
quantity $\tilde g_A^{(0)}$ in eq. (1) can be written through the
scale dependent axial constant $g_A^{(0)}$ \cite {RGINV}

$$
\tilde g_A^{(0)}=e^{-\int^{\alpha_s(\mu)}_0{\gamma(x)
\over \beta (x)}dx}g_A^{(0)},
$$
where $\gamma(x)$ is the singlet axial current anomalous dimension known at
the three-loop approximation \cite {andim} and
$\beta(x)$ is the QCD $\beta$--function determined at this order too
\cite {BF}.  Let us denote the first and second addendum in the
r.h.s. of eq.(1) as $a^{NS}$ and $a^{S}$, respectively. As a result,
Ellis-Jaffe sum rule will take the form $$
\int_0^1
g_1^p(x,Q^2)dx=a^{NS}(1,\alpha_s(Q^2))+a^{S}(1,\alpha_s(Q^2)). \eqno (3a)
$$
In these notations singlet part of the Ellis-Jaffe sum rule
in the lea\-ding order appro\-xi\-ma\-ti\-on can be written in the
form $$ a^{S}=(1-0.3333({\alpha_s(Q^2)\over\pi})){1\over9}\tilde
g_A^{(0)}. \eqno (3b) $$
Substituting the numerical values for the integral in the l.h.s., for
the coefficient function $C^{NS}(Q^2)$ \cite {NSCF} and also for the
nonsinglet axial constants \cite {NSAC}, one obtains a surprisingly
small value for the singlet part $a^{S}\simeq 0.02\pm0.01$. It is easy
to see that the perturbative corrections presented in (3a) do not save
the situation and the only possibility is to assume (in contrast with
the naive expectation) that the proton singlet axial constant $g_A^0$
itself is a small quantity. In what follows we will try to estimate
the magnitude for $g_A^0$.

Let us begin with the consideration of the anomaly equation for the flavor
singlet axial current with three active flavors
$$
\partial^\mu A_\mu^{(0)}=\sum_{q=u,d,s}2im_q{\bar q}\gamma_5 q +
{\alpha_s N_f\over 4\pi}G\tilde G. \eqno (4)
$$
Brief mention should be made of an important role of the mass terms in this
expression. At first glance it would seem that these terms are
negligibly small due to the smallness of the corresponding quark
masses.  Mass terms are often dropped out in the chiral limit.
However, it has been demonstrated in \cite {Brown} \cite {Johan} \cite
{Huang} that due to the fermionic zero modes over the instanton
background the matrix elements of the operators like ${\bar q}\gamma_5 q$
have singular behavior in the limit of zero quark mass. So, only the
whole quantity $m_q{\bar q}\gamma_5 q$ has a strict sense in the chiral
limit; moreover, this combination gives a nonzero contribution
when the quark masses are set to be equal to zero after all calculations are
performed.

On the other hand, one can assume that the mass term is absent in
the Lagrangian from the very beginning. If so, the derivative of the
axial current would be expressed through the gluon operator only.
However, this is the case when the QCD Lagrangian is chiral invariant
and consequently the corresponding generating functional could not
be able to produce a nonzero value for the quark condensate (if the
ordinary Feynman boundary conditions are assumed). The fulfillment of
the axial Ward identities in this case becomes doubtful too
(for the detailed discussion see ref. \cite {Huang}).
Another way of putting it is that one must calculate
quasiaverages instead of simple averages \cite {Bog} going to the limit
$m_q\rightarrow 0$ after this \cite {AFT}. Such a procedure allows one to
take into account consequences of the spontaneous breaking
of the chiral symmetry in the pure QCD. It has been demonstrated in
\cite {Johan} that the contributions coming from the mass terms in the
anomaly equation lead one to the nonrenormalization of
the $\theta$--term in QCD.  The mass terms play an important role in
our calculation too.

Now we are in a position to consider the matrix element of the  singlet axial
current over the proton states
$$
<P(k_1)|A_\mu^{(0)}|P(k_2)>= g_A^{(0)}(q^2) {\bar U(k_1)}\gamma_\mu
\gamma_5U(k_2)-iq_\mu g_P^{(0)}(q^2) {\bar U(k_1)}\gamma_5 U(k_2), \eqno (5)
$$
where $q_\mu=k_{1\mu}-k_{2\mu},$ $U(k) $ is the proton spinor.
It is well known that there are no massless excitations in the flavor
singlet axial channel even in the chiral limit \cite {U(1)}.
Consequently, the phenomenological expression for the pseudoscalar
constant $g^0_P$ can be written in the form $$
g^0_P=\sum_n {B_n \over q^2-m_n^2}, \eqno (6) $$
where $m_n^2$ are the mass squares of the co\-rres\-pon\-ding
pseu\-do\-sca\-lar particles in the flavor singlet case. These
quantities are not equal to zero even in the chiral limit.  This is an
important point where the singlet and nonsinglet constant calculations
drastically differ from each other. An analogous expression for the
nonsinglet pseudo\-sca\-lar constant could be presented as a sum of
terms containing poles in $q^2$ and the terms containing the
multiplier of an order of $o(m^2_q)$; as a result, the mixing between
the axial and the pseudoscalar constants will take place within the
corresponding sum rules. However, as we will see later, this is not
true for the flavor singlet case. Reason is that there are no poles in
eq. (6) even in the chiral limit.  The coefficients $B_n$ in this
expression are determined by the proton-meson interaction vertices.
Using now eq. (4), the matrix element (5) takes the form

$$
<P(k_1)|\partial^{\mu} A_\mu^{(0)}|P(k_2)>=(-i g_A^{(0)}(q^2)2m_p +q^2
g_P^{(0)}(q^2)) {\bar U(k_1)}\gamma_5 U(k_2), \eqno (7)
$$
where $m_p$ is the proton mass. Our aim will be to estimate
the value for $g_A^{(0)}$ by means of the QCD sum rule method \cite {SVZ}.
Following the ideology of this approach consider the three-point
correlation function $$
T_\mu^{\alpha\beta}(k_1,k_2)=\int e^{ik_1x-ik_2y}
<0|T^*\eta^{\alpha}(x)A_\mu^{(0)}(0){\bar\eta}^{\beta}(y)|0>dxdy=
$$
$$
=\int e^{ipx+iqy}
<0|T^*\eta^{\alpha}(x/2)A_\mu^{(0)}(y){\bar\eta}^{\beta}(-x/2)|0>dxdy,
\eqno (8)
$$
where $p=(k_1+k_2)/2,$ and $\eta^{\alpha}(x)$ is an interpolating current
for the proton state \cite {Ioffe}
$$
<0|\eta^{\alpha}(x)|P(k_1)>=\lambda_p U^{\alpha}(k_1).
$$
The phenomenological expression for this correlator can be written in
the form
$$
T_\mu^{\alpha\beta}(k_1,k_2)={ <0|\eta^{\alpha}|P(k_1)>
<P(k_1)|A_\mu^{(0)}|P(k_2)><P(k_2)|{\bar\eta^{\beta}}|0>\over
(k^2-m_p^2)^2}+...\eqno (9)
$$
where $k^2=k_1^2=k_2^2=p^2+q^2/4$ and the kinematical
condition (pq)=0  is assumed for simplicity. Dots
denote here the contributions with the one-pole term and the terms
without poles in $(k^2-m_p^2) $. Multiplying expression (9) over
$iq^{\mu}$ and using (7) we obtain $$
iq^{\mu}T_\mu^{\alpha\beta}(k_1,k_2)={i\lambda_p^2 g_A^0(q^2)2m_p^2
\hat q\gamma_5\over (k^2-m_p^2)^2}+
$$
$$
{i\lambda_p^2 g_A^0(q^2)q^2[{\hat q}/4+m_p]\gamma_5+\lambda_p^2q^2
g^0_P(q^2)({\hat k}_1-m_p)\gamma_5({\hat k}_2-m_p)\over (k^2-m_p^2)^2}+...
\eqno (10)
$$
Let us consider now
the operator product expansion (OPE)
for our three-point correlation function. (Hereafter, we will follow
the definitions of \cite {INR}). The bilocal OPE for the correlator
under investigation looks like $$
iq^{\mu}T_\mu^{\alpha\beta}(k_1,k_2)|_{|p^2|\rightarrow\infty} =\int
e^{ipx+iqy} <0|T^*\eta^{\alpha}(x/2)\partial^\mu
A_\mu^{(0)}(y){\bar\eta}^{\beta}(-x/2)|0>dxdy|_{|p^2|\rightarrow\infty}
$$
$$
=\sum_n C_n(p)i\int e^{iqy}
<0|T^*{\hat O}_n(0)\partial^\mu
A_\mu^{(0)}(y)|0>dy+\sum_n R_n(p,q)<0|{\hat O}_n(0)|0>.  \eqno (11)
$$
The first sum in the r.h.s. of this expansion is determined by the
vacuum expectation
values of some bilocal operators and the second one gives the
contributions of the local operator vacuum expectation values.
The operators ${\hat O_n(0)}$ are generated within the OPE for the
proton interpolating currents $$ i\int e^{ipx} T^*\eta^{\alpha}(x/2)
{\bar\eta}^{\beta}(-x/2)dx= \sum_n C_n(p){\hat O}_n(0) $$ It is worthwhile to
point out here that the bilocal VEVs play the crucial role in this approach,
in fact, they give the dominant contribution in the case of the sum
rules for the proton pseudoscalar constants (for a detailed discussion
see ref. \cite {INR}).  However, we will demonstrate that the bilocal
VEVs cancel each other in the singlet case and only the local part
of the OPE determines the value for the proton singlet axial constant.
Let us turn to the consideration of the
correlators presented in the  r.h.s. of eq. (11) $$ P_n(q^2)=i\int
e^{iqy}<0|T^*{\hat O}_n(0)\partial^\mu A_\mu^{(0)}(y)|0>dy.
$$
The dispersion
relation for this quantity looks like
$$
P_n(Q^2=-q^2)={1\over\pi}\int {ImP_n(s)ds\over
s+Q^2}+subtractions.
$$
Some remarks concerning the peculiarities of this expression are in order
here. It is well known that the correlator for two singlet axial currents
$<A_\mu^{(0)}  A_\nu^{(0)}>$ contains the so-called Kogut-Susskind pole
\cite {KS} beyond the intermediate physical particle contributions.
This pole term is produced by the collective excitations over the complicated
vacuum \cite {Collect}. Within the momentum representation for
$<A_\mu^{(0)}  A_\nu^{(0)}>$  the Kogut-Susskind pole gives an
ad\-di\-ti\-onal term pro\-por\-ti\-onal to $(q_\mu q_\nu)/q^2$. Due
to singularities in $q^2$, the existence of such a contribution makes
it difficult to use the sum rule formalism. However, in our case,
after multiplication of the corresponding correlator over  $q^\mu$ the
Kogut-Susskind contribution becomes the polynomial in mo\-men\-ta and,
consequently, could be absorbed in the subtractional part  of the
dispersion relation. Then, using the Borel transformation this
polynomial disappears from the sum rules under investigation. Later
on, we shall follow this way.

On the other hand,
the phenomenological expression for the imaginary part of
the correlator under consideration looks like $$
ImP_n(s)=\pi\Delta(s)<0|{\hat O}_n|0><0|\sum_{q=u,d,s}2im_q{\bar
q}\gamma_5 q + {\alpha_s N_f\over 4\pi}G\tilde G|0>+
$$ $$ +\pi\sum_k\delta(s-m_k^2)<0|{\hat
O}_n|P_k><P_k|\sum_{q=u,d,s}2im_q{\bar q}\gamma_5 q + {\alpha_s
N_f\over 4\pi}G\tilde G|0>,  $$
where $\Delta(s)$ determines the complicated vacuum contribution.
The second addendum in the r.h.s.  of this relation gives the
contribution only to the proton pseudoscalar constant $g^0_P$ and it
is the first term in the r.h.s. which is responsible for $g^0_A$.
However, following \cite {Brown} \cite {Huang} this last is equal to
zero.  Indeed, let us integrate over the fermionic fields in the path
integral representation for $<0|{\bar q}\gamma_5 q|0>$ with
the subsequent calculation of the corresponding determinant in terms of
the gluon fields. As an answer one can obtain \cite {Brown} \cite
{Huang} $$ <0|{\bar q}\gamma_5 q|0>=-{\alpha_s\over i 8\pi }{G\tilde
G\over m_q}.  $$ Consequently $$ <0|\sum_{q=u,d,s}2im_q{\bar
q}\gamma_5 q + {\alpha_s N_f\over 4\pi}G\tilde G|0>=0.  $$
We can conclude that the bilocal operator VEVs do not affect  the
sum rule for the proton singlet axial constant. It is important to point out
that this statement is based  on the assumption of
absence of the massless excitations in the singlet channel in the
chiral limit (relation with the $U(1)$ problem, see ref. \cite {EST}).
Now it is evident that  the proton singlet axial constant must be
small in comparison with the nonsinglet one. The point is that the
local part of the OPE determining the value for $g_A^0$ gives the
contributions of the order $\alpha_s^2$ only.  Let us turn to the
consideration of these terms.  For practical calculations of the local
part it is convenient to introduce the following quantity $$
D_\tau^{\alpha\beta}=[{\partial\over {\partial}q^\tau}
(iq^{\mu}T_\mu^{\alpha\beta})]|_{q=0}= $$ $$ ={i\lambda_p^2
g_A^0(0)2m_p^2 \gamma_\tau\gamma_5\over (p^2-m_p^2)^2}+
{C\gamma_\tau\gamma_5\over(p^2-m_p^2)}+...,\eqno (12) $$ where $C$ is
some unknown constant.  The leading contributions to the sum rule for
$g_A^0$ come from the three-loop perturbative diagram and the one-loop
diagram with the four-fermion condensate. Our main assumption is that
the perturbative three-loop contribution is suppressed by the
loop-factor $1/(4\pi^2)^3$ and could be neglected in the case of rough
estimation. The only thing we have done is the calculation of the
four-fermion condensate contributions. In such an approximation, the
corresponding sum rule takes the form $$
{\lambda_p^2g^0_A(0)2m^2_p\over
(P^2+m_p^2)^2}+{iC\over(P^2+m_p^2)}+...= C^2_F({\alpha_s\over
\pi})^2{1\over 2}{<{\bar u}u>^2\over P^2}ln(P^2/{{\mu}^2}), $$ where
$C_F $ is the Casimir operator of the fundamental representation of
$SU(3)_c$ group ($C_F=4/3$) and $P^2=-p^2$.  Multiplying this
expression over the quantity $P^2+m_p^2$ and using then the Borel
transformation, we obtain the following sum rule for the proton singlet
axial constant $$ 2\lambda_p^2g^0_A(0)e^{-{1\over \tau}}\simeq
C^2_F({\alpha_s\over \pi})^2 {1\over 2}<{\bar u}u>^2(\tau^2-\tau),
\eqno (13) $$
where $\tau= M^2/m_p^2$ and $M^2$ is the Borel parameter.
Substituting the known values for the quark condensate and all
parameters in eq. (13) \cite {Ioffe} we obtain the following result

$$
g^0_A(0)|_{\mu=1GeV}e^{-{1\over \tau}}\simeq 0.025(\tau^2-\tau).
$$
The relative stability in these sum rules is reached
when $\tau\simeq 1.3-1.5$ and consequently
$$
g^0_A(0)|_{\mu=1GeV}\simeq 0.02-0.03.
$$
So our crude estimation leads to the value for the proton singlet axial
constant which is sufficiently small in comparison with unity and the
nonsinglet constants.

Now, using the obtained value for the proton singlet axial constant one can
estimate the $s$ -- quark sea contribution to the proton matrix element.
Indeed, writing the relation  $g^{(0)}_A-g^{(8)}_A=3\Delta s$ and
substituting the known value for $g^{(8)}_A$ \cite {NSAC}, one can obtain
that $\Delta s|_{\mu=1GeV}\simeq -0.16$.  This estimate is in
qualitative agreement with the value obtained from the precision fit
of the experimental data \cite {EK}.

In conclusion, let us make remark concerning our consideration.
The result obtained
here is qualitative in its essence and could be considered  as some
crude approximation only.  For a more precise calculation of the
value for the proton singlet axial constant it is  necessary to take
into account both the three-loop perturbative contribution and the
dimension-seven quark-gluon operator contribution within the QCD sum
rule approach. However, one could not expect that these contributions
will lead to sufficient increasing in the value for the proton singlet
axial constant. Missing contributions are of an order of $\alpha_s^2$
only.  We have demonstrated that correct treatment of the anomalous
contributions leads to sufficient numerical suppression for the value of
the proton singlet axial constant (in terms of sum rules the
cancellation of the bilocal contributions takes place). This statement is
based on the assumption that there are no massless excitations in the flavor
singlet channel even in the chiral limit.

So, we have estimated the value for the proton singlet axial constant.
This value is sufficiently suppressed in comparison with the naively
expected one. The value of the $s$ -- quark sea contribution for the proton
matrix element is also estimated.

The author is grateful to S.A. Larin, A.A. Pivovarov and A.N.
Tavkhelidze for conversations. A.L. Kataev is acknowledged
for useful suggestions. The work is
partly supported by the  Russian Fund of the Fundamental Research, Grants N
94-02-04548a, N 94-02-14428 and "Bazis Bank" graduate student fellowship.

\end{document}